# Towards Mapping and Assessing Sidewalk Accessibility Across Socio-cultural and Geographic Contexts

**Jon E. Froehlich**
Computer Science
University of Washington
jonf@cs.uw.edu

**Michael Saugstad**
Computer Science
University of Washington
saugstad@cs.uw.edu

**Manaswi Saha**
Computer Science
University of Washington
manaswi@cs.uw.edu

**Matthew Johnson**
Computer Science
University of Washington
matt8@cs.uw.edu

**Abstract**
Despite the important role of sidewalks in supporting mobility, accessibility, and public health, there is a lack of high-quality datasets and corresponding analyses on sidewalk existence and condition. Our work explores a twofold vision: first, to develop scalable mechanisms to locate and assess sidewalks in cities across the world, and second, to use this data to support new urban analyses and mobility tools. We report on two preliminary urban science explorations enabled by our approach: exploring geo-spatial patterns and key correlates of sidewalk accessibility and examining differences in sidewalk infrastructure across regions.

*Accepted to Data4Good.*

**Introduction**
Sidewalks are a unique form of public infrastructure: they provide a safe, off-road pathway for pedestrians, help interconnect mass transportation services like bus and rail, and support commerce and recreation [4,8]. For individuals with a mobility disability, sidewalks play a crucial role in independence [12], quality of life [9], and overall physical activity [3]. Despite their importance, sidewalks are often relegated to secondary (or even non-existent) positions compared to auto-centric street networks in modern mapping tools offered by Google, Apple, and OpenStreetMaps. Compared to their road counterparts, a key challenge with assessing sidewalk existence and quality is the systematic lack of datasets, open standards, and analysis tools [5]. This data gap is only widening as car companies and big tech increasingly attach high-resolution sensors to vehicles to collect data for autonomous (or semi-autonomous) driving.

The lack of sidewalk data impacts not just how algorithms route us through cities but also how transportation and urban planners study mobility and urban design. As street datasets grow, researchers are developing new computational and visualization techniques for modeling and comparing urban morphology and street topology across cities [1,2]. But this work does not incorporate sidewalks. Moreover, although sidewalks are increasingly presented as offering public health, economic, environmental, and accessibility benefits [7,11], few studies have examined sidewalks in developing countries. Those that do [15,17] often cast "developing countries" as a single amalgam rather than with a specific regional and cultural lens. Again, we argue, a key problem is data.

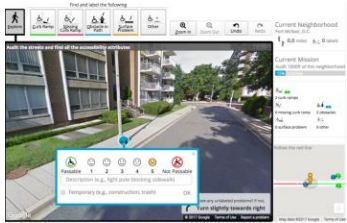

**Fig. 1:** The Project Sidewalk interface. Here, a user selects the blue "Obstacle" label type for the pole sidewalk obstruction and rates it a '5' as an impassable barrier for wheelchair pedestrians.

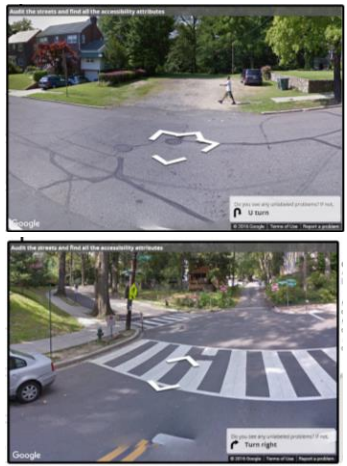

**Fig. 2:** Two example issues found in Project Sidewalk: (top) A pedestrian walking on grass after a sidewalk ends and (bottom) a newly painted crosswalk to an intersection without a curb ramp.

In our work, we are developing new sidewalk data collection and analysis techniques using a combination of remote crowdsourcing, machine learning (ML), and online map imagery (*e.g.,* satellite images, street panoramas). Our overarching aim is twofold: first, to develop scalable hybrid human+ML techniques capable of mapping and assessing every sidewalk in the world and second, to leverage this data to enable new pedestrian-oriented mapping tools (*e.g., AccessScore* [10]), to provide increased transparency and accountability about city accessibility, and to support new urban analytics pursuits not previously possible.

Our most recent effort, called *Project Sidewalk*, enables online crowdworkers to remotely label sidewalks and find and identify accessibility problems by *virtually* walking through city streets in Google Street View (Fig. 1). To train, engage, and sustain users, we apply basic game design principles such as interactive onboarding, mission-based tasks, and progress dashboards. Labels are used to develop new data analytic and visualization tools to highlight and explore (in)accessible areas of a city (*e.g.,* LabelMap) but also to train deep learning networks to assess sidewalks *automatically*—further scaling our approach [16]. In a pilot deployment in Washington, DC [14], Project Sidewalk users virtually audited 3,000+ miles of streets and labeled over 255,000 sidewalk accessibility problems with 92% accuracy.

Building on our successful pilot deployment and working with local and international partners such as *Open Columbus* and *Liga Peatonal*, we have now deployed Project Sidewalk in four additional cities, including Seattle, WA, Newberg, OR, Columbus, OH, and Mexico City, MX, with two additional cities launching soon. While there is a range of interesting technical, system-level research questions to address, such as how to optimally route crowdworkers through cities to find and assess sidewalks and how to combine computer vision with manual work to maximize efficacy, we are particularly interested in exploring new data science-oriented questions enabled by our growing datasets, such as:

- What are the geo-spatial patterns and key correlates of sidewalk accessibility? How does accessible infrastructure correspond to racial and socio-economic factors? How do these patterns compare across cities?
- What do sidewalks look like across cities and countries? How do their designs vary with respect to accessibility? What region-specific accessibility barriers exist and how do these barriers reflect the socio-cultural and-political context of that region?

Below, we describe our initial work in this area and close with a discussion of challenges and future work.

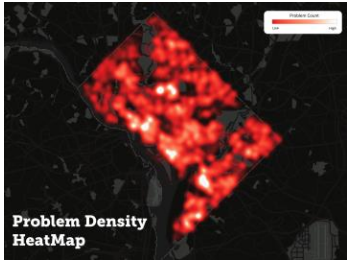

**Fig. 3:** A heatmap of sidewalk accessibility problems found with Project Sidewalk in DC.

| | Users | Total km Complete | Total Labels |
|---|---|---|---|
| **Seattle, WA** | 1,576 | 1,608.3 | 92,127 |
| **Columbus, OH** | 263 | 185.7 | 16,277 |
| **Newberg, OR** | 203 | 224.6 | 16,076 |
| **Azcapotzalco, MX** | 222 | 80.2 | 5,864 |

**Table 1:** The number of users, total km audited, and sidewalk accessibility labels collected across our four newest Project Sidewalk deployment cities.

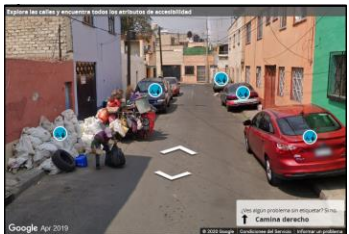

**Fig. 4:** A series of sidewalk obstacles in Azcapotzalco, MX found in Project Sidewalk and tweeted by one of our users.

### What are the geo-spatial patterns and correlates of sidewalk accessibility?

While there is significant research exploring environmental and neighborhood correlates of walkability and transportation usage (*e.g.,* [3,5]), we could find no prior work that specifically incorporates sidewalk availability, quality, and accessibility as factors in their analyses. Using Project Sidewalk's dataset, we aim to examine how sidewalk quality correlates with socio-demographic, land use, and economic factors.

As a start, we conducted an informal visual analysis of our Washington, DC dataset using heatmaps (Fig. 3) and found a higher density of surface problems and sidewalk obstacles along the southeastern corridor of the city along the Anacostia River, a historically Black neighborhood. Interestingly, we also found a higher precedence of accessibility problems in one of the most affluent DC neighborhoods, Georgetown, perhaps because of policies aimed at preserving historic cobblestone walkways—but at a cost of accessibility. More work is needed to formalize these analyses, compare and contrast them across cities, and develop and publish analytic tools that help others interactively identify new geo-spatial patterns and disparities.

### How does sidewalk infrastructure compare across cities and countries?

As we deploy Project Sidewalk in new regions, we are interested in exploring how sidewalk infrastructure differs across regions and to uncover underlying social, historical, political, and economic influences. As one example: in 2019, we were contacted by Liga Peatonal ("Pedestrian League"), a Mexico-based NGO focused on pedestrian improvements to increase the safety and accessibility of public spaces in Mexico. Working closely with their staff, we translated Project Sidewalk's interfaces into Spanish, added locale-specific label tags, and created co-branded logos and advertising. As a pilot, we deployed into a single Mexico City municipality called Azcapotzalco, which has a population of over 400,000 people across 33.6 km$^2$.

While our deployments are ongoing, we describe some initial quantitative and qualitative findings between Azcapotzalco sidewalks and three of our newest deployment cities in the US (Table 1). With Project Sidewalk, users find, label, and rate the severity (1 to 5 where 5 is worse) of sidewalk accessibility issues, including curb ramps, surface problems, and physical obstacles. In a preliminary analysis, we found that identified issues in Azcapotzalco were rated worse, on average, than labels of the same type in the other three cities (Table 2; Fig. 4). For example, the average curb ramp was rated as a 2.8 severity *vs.* 1.5 in the other cities (higher is worse). To better understand this difference, we have begun qualitatively examining applied labels and identified a pattern of degraded and/or poorly designed curb ramps (Fig 5).

| | Curb Ramp | Missing C. Ramp | Missing Sidewlk | Obstacle | Surface Problem |
|---|---|---|---|---|---|
| **Seattle, WA** | 1.5 (0.7) | 3.8 (1.0) | 4.0 (0.8) | 3.2 (1.1) | 2.9 (0.9) |
| **Columbus, OH** | 1.4 (0.7) | 3.9 (1.1) | 4.3 (1.1) | 2.2 (1.4) | 2.1 (1.0) |
| **Newberg, OR** | 1.5 (0.7) | 3.9 (1.0) | 3.9 (0.9) | 3.1 (1.1) | 2.7 (1.0) |
| **Azcapot., MX** | 2.8 (1.4) | 4.7 (0.6) | 4.6 (0.7) | 4.1 (0.9) | 3.6 (1.1) |

**Table 2:** Severity ratings of five sidewalk label types in Project Sidewalk across four deployment cities. Ratings are 1-5 where 5 is worst. Standard deviation is in parentheses.

Sidewalk accessibility issues in Azcapotzalco are not only rated more severely, they are also more frequent. When examining the number of labels per 100m in each city, we found that Azcapotzalco has comparatively fewer curb ramps and significantly more sidewalk obstacles and surface problems (Table 2). Additional work is needed to understand these differences, including interviews with our local partners to ascertain potential policy and cultural influences.

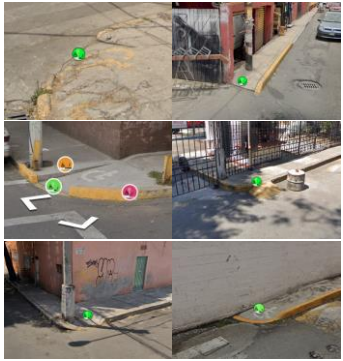

**Fig. 5:** Example curb ramps in Azcapotzalco, MX rated as poor quality.

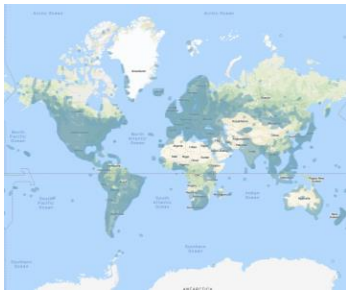

**Fig. 5:** Google Street View is in 83 countries but notably lacking in northern and central Africa, large parts of Asia, and Russia. Light blue represents partial coverage; dark blue is full coverage. From [6].

|  | Curb Ramp | Missing C. Ramp | Missing Sidewalk | Obstacle | Surface Problem |
|---|---|---|---|---|---|
| **Seattle, WA** | 2.1 | 1.1 | 1.5 | 0.3 | 0.6 |
| **Columbus, OH** | 4.3 | 0.3 | 1.4 | 1.2 | 1.4 |
| **Newberg, OR** | 1.9 | 0.9 | 3.0 | 0.4 | 0.9 |
| **Azcapotzalco, MX** | **1.0** | **1.2** | 0.3 | **1.8** | **2.5** |

**Table 3:** Frequency of sidewalk accessibility labels (per 100 meters) across four deployment cities. For curb ramps, higher is better. For other columns, lower is better.

## Discussion

In this workshop paper, we reported on our initial attempts to explore disparities in sidewalk infrastructure within and across cities using Project Sidewalk data. We briefly reflect on current challenges and future plans.

**Who does the work?** Project Sidewalk is a crowdsourcing tool that largely relies on volunteer labor for sidewalk labeling and assessment (although we have also run experiments with Mechanical Turk). As a remote tool, anyone with an Internet connection and a computer or smartphone can contribute—we have had users in Europe audit sidewalks in Azcapotzalco, for example. While far more scalable and inexpensive than traditional on-the-ground walkability audits, Project Sidewalk users may lack local knowledge and cultural awareness when contributing data outside their own region. Moreover, the reliance on technology itself excludes potential users. Liga Peatonal recently asked us to print out paper "audit" forms for some members in their community (to be manually filled out and entered into the Project Sidewalk database).

**Google Street View.** While our techniques should work with any streetscape imagery dataset, including Mapillary, CycloMedia, and Bing Streetside, Project Sidewalk is currently dependent on Google Street View. In 2017, Google announced that it had captured more than 10 million miles (16 million km) of Street View imagery across 83 countries [6]; however, there is a notable lack of coverage in northern and central Africa, large parts of Asia, and Russia (Fig. 5)—which is an unfortunate limitation and its own source of bias. Relatedly, the recency and frequency with which Google Street View cars collect data is important and may also be biased with socio-economic factors. Like Mapillary, Google now allows end-users to submit their own street images, which may partially mitigate this problem but assumes technology access and literacy.

**Affordances and limitations of Project Sidewalk data.** Project Sidewalk image-based label data affords both quantitative analyses—*e.g.,* what areas of a city have poor sidewalk infrastructure—and more limitedly, qualitative analyses of identified problems as well as sidewalk usage patterns captured by the streetscape images (*e.g.,* location of vending stands, café seating); however, addressing and ultimately improving sidewalk infrastructure is a complex socio-political and economic problem [13]. Moreover, the ways in which sidewalks are designed, used, and maintained fundamentally reflects the values, culture, and economics of a region and people. To fully assess sidewalk infrastructure, a mixed-methods approach is necessary, involving ethnographic observation of sidewalk usage, interviews and surveys of local populations and stakeholders, and studying local legislation and policy.

**Future cities.** Finally, and building on the above points, as we deploy Project Sidewalk into additional cities, it is essential that we work with local partners both in our design and analysis to better understand and support region-specific contexts.

## Acknowledgements

Project Sidewalk is a large, ongoing effort that has benefited from the contributions of many. We thank our partners at Liga Peatonal for supporting our Mexico-based deployments, our undergraduate research assistants, and the NSF (#1302338), Google, the Sloan Foundation, and the Pacific Northwest Transportation Consortium for partially funding this work.